\documentclass[aip,pra,twocolumn,showpacs]{revtex4}

\usepackage{graphicx,color}
\usepackage{amsmath}
\usepackage{longtable}
\begin{document}

\title{FCC-BCC-Fluid triple point for model pair interactions with variable softness}

\author{Sergey A. Khrapak\footnote{Also at Joint Institute for High Temperatures, 125412 Moscow, Russia; Electronic mail: skhrapak@mpe.mpg.de} and Gregor E. Morfill}

\affiliation{Max-Planck-Institut f\"ur extraterrestrische Physik, D-85741 Garching, Germany}

\date{\today}

\begin{abstract}
It is demonstrated that the coordinates of the fcc-bcc-fluid triple point of various model systems are located in a relatively narrow region, when expressed in terms of the two proper variables, characterizing the softness and strength of the interaction force at the mean interparticle separation. This can be regarded as a consequence of the ``corresponding states principle'' for strongly interacting particle systems we have put forward recently [S. A. Khrapak, M. Chaudhuri, and G. E. Morfill, J. Chem. Phys. {\bf 134}, 241101 (2011)]. The related possibilities to predict the existence and approximate location of the fcc-bcc-fluid triple point for a wide range of pair interactions with variable softness are illustrated. Relation of the obtained results to experimental studies of complex (dusty) plasmas are briefly discussed.
\end{abstract}


\maketitle

Many systems of strongly interacting particles in thermodynamic equilibrium are known to form face-centered-cubic (fcc) or body-centered-cubic (bcc) lattices in the solid phase. Normally, for sufficiently soft interactions there is a tendency to form a bcc solid. On the contrary, steep interactions favor the fcc structure. When the interaction potential exhibits variable softness, a polymorphic phase transition between fcc and bcc solids can be present and the fcc-bcc-fluid triple point emerges on the respective phase diagram.

Finding the location of fcc-bcc phase transition remains a difficult and resource demanding (computational) task since usually a very tiny difference between the free energies of the corresponding phases has to be resolved. In this Letter we discuss a very simple approach to {\it approximately} locate the fcc-bcc-fluid triple point. The approach is based on analyzing the shape of the pair interaction potential only and does not require any kind of numerical simulations.

The qualitative similarities in the phase behavior of model systems with soft repulsive interactions have been discussed previously from various perspectives. In particular, Prestipino {\it et al}.~\cite{PrestipinoPRE,PrestipinoJCP} discussed the common topology of the phase diagrams of the Gaussian core model (GCM), inverse-power-law (IPL), and Yukawa model potentials in the vicinity of the fcc-bcc-fluid triple point. In an attempt to relate the phase behaviors of the GCM and Yukawa potentials to that of the IPL model, they required the logarithmic derivatives of the three potentials to match, for separations close to the average nearest-neighbor distance. As a result, the effective exponents for the GCM and Yukawa interactions at the triple point were shown to lie not too far from the corresponding IPL value. This was a useful observation, although perhaps insufficient for direct practical applications.

As we conjectured recently~\cite{Principle}, similarities in the behavior of two different systems of strongly interacting particles can be expected when the interparticle force and its first derivative (evaluated at the mean interparticle distance) in one system are equal to those in the other system. The simplistic arguments behind this conjecture are as follows: In a {\it strongly coupled} state the particles form a regular structure where large deviations of the interparticle separation from its average value are very seldom. Consequently, the state of the system should be virtually insensitive to the exact shape of the interaction potential at short distances. If, in addition, the potential decays sufficiently fast for distances beyond the mean interparticle separation, its long-range asymptote is of little importance for the thermodynamics of the system~\cite{Note1}. The behavior of the potential at the average interparticle separation plays a dominant role. Physically, the force and its first derivative are the two quantities which should mainly matter. Using the IPL potential as a reference system, we identified the two scaled variables which conveniently characterize the state of a strongly coupled system of particles interacting via the pair potential $U(r)$. These are the {\it generalized softness parameter} $s=\left[-1-U''(\Delta)\Delta/U'(\Delta)\right]^{-1}$ and the {\it reduced force} ${\mathcal F}=-U'(\Delta)\Delta/T$, where $\Delta$ is the mean interparticle separation and $T$ is the temperature (in energy units). The approximate ``corresponding states principle'' postulates that two different systems of strongly interacting particles having the same values of $s$ and ${\mathcal F}$ behave alike~\cite{Principle}. As a test of the {\it vitality} of this principle, we demonstrated that the melting curves of various model systems such as IPL, Yukawa, conventional Lennard-Jones (LJ), $n-6$ LJ, $\exp -6$ and (slightly less successfully) GCM are essentially collapsing on a single curve when plotted in the $(s, {\mathcal F})$ plane and, hence, the ``universal melting curve'' emerges~\cite{Principle}.

The question we address here is whether this corresponding states principle can also be useful to estimate (at least approximately) the location of the fcc-bcc-fluid triple point. In order to answer it, we carefully analyze the available data on the fcc-bcc phase transition for three selected repulsive potentials. We show that the triple point coordinates in the $(s, {\mathcal F})$ plane are indeed relatively close for all the systems considered. In addition, the phase diagrams in the vicinity of the triple point are topologically equivalent.

We consider the IPL potential, $U(r)=\epsilon(\sigma/r)^n$, the Yukawa (exponentially screened Coulomb) potential, $U(r)=\epsilon(\sigma/r)\exp(-r/\sigma)$, and the GCM potential, $U(r)=\epsilon\exp(-r^2/\sigma^2)$. Here $\epsilon$  and $\sigma$ are the energy and length scales, and $r$ is the distance between the two interacting particles. The IPL potential defines a system of ``soft'' spheres which can serve as a model for a wide range of physical systems, including the one-component plasma (OCP) and simple metals under extreme thermodynamic conditions. At
$n\rightarrow\infty$ it approaches the hard sphere (HS) limit. The Yukawa (Dedbye-H\"{u}ckel) potential is extensively used in the context of colloidal suspension, physics of plasmas, and complex (dusty) plasmas~\cite{FortovPR,Book,ChaudhuriSM,Likos,Malescio}. The GCM potential~\cite{Stillinger} is frequently used in the context of soft matter physics, e.g. to describe effective interactions in many body systems of polymer solutions~\cite{Likos,Flory,Louis}

The phase state of the considered systems in thermodynamic equilibrium can be conveniently characterized by the reduced density $\rho_*=\rho\sigma^3$ and temperature $T_*=T/\epsilon$ (For the IPL potential, the single variable $\rho_*T_*^{-3/n}$ fully describes the thermodynamic state of the system). Another possible choice of state variables for purely repulsive interactions uses the {\it coupling parameter} $\Gamma=U(\Delta)/T$ and the {\it screening parameter} $\kappa=\Delta/\sigma$, where $\Delta=\rho^{-1/3}$ is the structure-independent interparticle distance. This choice is particularly suitable for Yukawa interactions and is commonly used in the field of complex (dusty) plasmas~\cite{FortovPR,Book}.

The phase portraits of the IPL, Yukawa, and GCM systems are relatively well studied~\cite{Hoover,Dubin,Agrawal,PrestipinoJCP,Kremer,Robbins,Meijer,Stevens,Dupont,Hamaguchi,Vaulina,Hoy,Stillinger1,Lang,PrestipinoPRE,KhrapakPRL}. Concerning the fcc-bcc-fluid triple point, Agrawal and Kofke~\cite{Agrawal} estimated its location for the IPL system as $\rho_*T_*^{-3/n}\simeq 2.173$ (in the fluid phase) and $1/n\simeq 0.16$. They noted, however, that this estimate is subject to significant uncertainty. Later, Prestipino {\it et al}.~\cite{PrestipinoJCP} performed more detailed study of the IPL phase diagram near the triple point. Their result is $\rho_*T_*^{-3/n}\simeq 1.787$ (in the fluid phase) and $1/n\simeq 0.14$. In the same work the triple point temperature of the GCM system has been estimated as $T_*\simeq 0.0031$. The corresponding fluid phase density is $\rho_*\simeq 0.093$. Dupont {\it et al}.~\cite{Dupont} reported the following triple point coordinate for the Yukawa model: $\kappa=6.75$ and $T_*\simeq 0.43\times 10^{-4}$. Later, Hamaguchi {\it et al}.~\cite{Hamaguchi} determined the location of the triple-point in Yukawa systems as $\kappa= 6.90$ and $\Gamma\simeq 3.50$. Finally, in a recent paper by Hoy and Robbins~\cite{Hoy}, the triple point of Yukawa systems has been located at $\kappa=7.70$ and $\Gamma\simeq 3.11$. For consistency, we take the fluid density at the triple point where appropriate. For the data related to the Yukawa potential, the triple-point density has a single value, defined by the estimated value of $\kappa$.

\begin{table}
\caption{\label{Tab1} The estimated values of the parameters $\Gamma$, ${\mathcal F}$ and $s$ at the fcc-bcc-fluid triple point for various model interaction potentials.}
\begin{ruledtabular}
\begin{tabular}{lllll}
Potential & $\Gamma_{\rm TP}$ & ${\mathcal F}_{\rm TP}$ & $s_{\rm TP}$ & Ref.  \\ \hline
GCM & 2.47 & 24.08 & 0.129 & \cite{PrestipinoJCP} \\
IPL & 3.98 & 28.46 & 0.14 & \cite{PrestipinoJCP} \\
IPL & 5.05 & 31.57 & 0.16 & \cite{Agrawal} \\
Yukawa & 4.03 & 31.26 & 0.145 & \cite{Dupont}  \\
Yukawa & 3.50 & 27.65 & 0.142 & \cite{Hamaguchi}  \\
Yukawa & 3.11 & 27.09 & 0.128 & \cite{Hoy} \\
\end{tabular}
\end{ruledtabular}
\end{table}

The corresponding values of $\Gamma_{\rm TP}$, ${\mathcal F}_{\rm TP}$, and $s_{\rm TP}$ at the triple point are summarized in Table~\ref{Tab1}. The scattering in the reported $\Gamma_{\rm TP}$ values is relatively large, up to a factor of two (see the first column in Table~\ref{Tab1}). On the other hand, the scattering in the triple-point values of ${\mathcal F}$ and $s$ are considerably less pronounced. The difference between the corresponding values for different interaction potentials is comparable to that from different studies of a single interaction (see e.g. the data for the Yukawa potential). Some (weak) dependence of $s_{\rm TP}$ and ${\mathcal F}_{\rm TP}$ on the shape of the interaction potential cannot be completely ruled out. At the same time, all the data points fall in the relatively narrow range $s_{\rm TP}\simeq 0.14\pm 0.02$ and ${\mathcal F}_{\rm TP}\simeq 28.0\pm 4.0$. This supports the ``corresponding states principle'' conjecture formulated in terms of the equality between the values of the parameters $s$ and ${\mathcal F}$, characterizing different interacting particle systems. In addition, the conditions
\begin{equation}\label{criterion}
s(\rho_*)\simeq 0.14,\qquad {\mathcal F}(\rho_*,T_*)\simeq 28,
\end{equation}
are expected to yield reasonable approximations for $\rho_*$ and $T_*$ at the fcc-bcc-fluid triple point of many other interacting particle systems (we will verify this towards the end of this Letter).

\begin{figure}[t!]
\centering
\includegraphics[width= 7.2 cm]{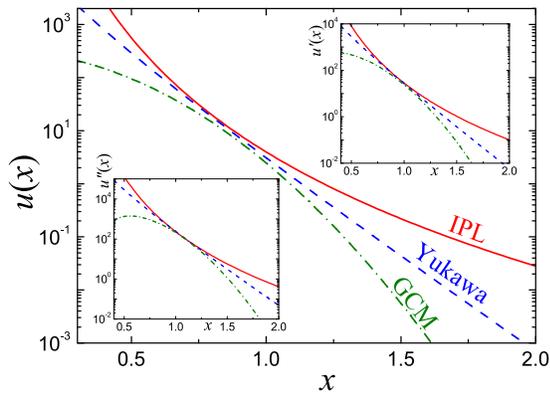}
\caption{(Color online). IPL, Yukawa, and GCM interaction potentials in the vicinity of the mean interparticle separation. Potentials are normalized by the temperature, distance is in units of the mean interparticle separation, $x=r/\Delta$. The parameters of the potential correspond to $s=0.14$ and ${\mathcal F}=28$ (see the text for details). The upper (lower) inset shows the first (second) derivative of the potentials.}\label{Fig1}
\end{figure}

In order to illustrate the relation between the corresponding states principle and the behavior of the interaction potentials around the mean interparticle separation, we have plotted the considered potentials in Fig~\ref{Fig1}. The potentials are normalized by the temperature, $u(x)=U(x)/T$, and the reduced distance is $x=r/\Delta$. The parameters of each potential are uniquely determined from the conditions $s=0.14$ and ${\mathcal F}=28$ (which are taken as characteristic for the fcc-bcc-fluid triple point). We observe that there is a non-negligible difference in the potentials around $x=1$, which is observable even on the logarithmic scale of Fig~\ref{Fig1}. At the same time, the first two derivatives are almost coinciding in some vicinity around $x=1$.

\begin{figure}
\centering
\includegraphics[width= 6.5 cm]{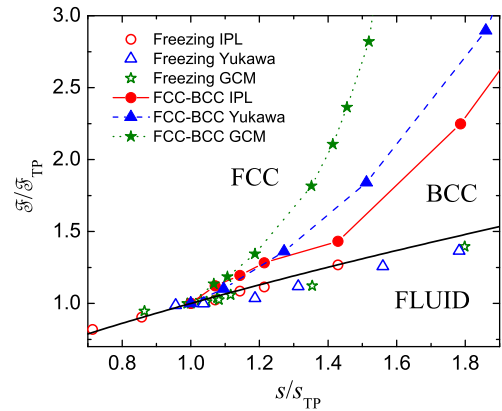}
\caption{(Color online). Phase boundaries of the IPL, Yukawa, and GCM systems near the fcc-bcc-fluid triple point in the plane of reduced parameters $s/s_{\rm TP}$ and ${\mathcal F}/{\mathcal F}_{\rm TP}$. Solid symbols correspond to the fcc-bcc phase transition, open symbols mark the fluid-solid (freezing) phase transition. The data for the IPL system (circles) are from Ref.~\cite{PrestipinoJCP} (freezing) and Refs.~\cite{Dubin,PrestipinoJCP} (fcc-bcc transition). The data for the Yukawa system (triangles) are from Refs.~\cite{Hamaguchi} (fluid-solid transition) and \cite{Hoy} (fcc-bcc transition). The data for the GCM system (stars) are from Ref.~\cite{PrestipinoJCP} (both the freezing and fcc-bcc transitions).}\label{Fig2}
\end{figure}

Next we use numerical results related to the fluid-solid and fcc-bcc phase transitions of the considered model systems and plot the corresponding phase boundaries in Fig.~\ref{Fig2}. In this figure
a portion of the unified phase diagram near the fcc-bcc-fluid triple point is shown in the plane of reduced variables $s/s_{\rm TP}$ and ${\mathcal F}/{\mathcal F}_{\rm TP}$. The numerical data are taken from Refs.~\cite{Dubin,PrestipinoJCP} (IPL), \cite{Hamaguchi,Hoy} (Yukawa), and \cite{PrestipinoJCP} (GCM), for details see caption of Fig.~\ref{Fig2}. To evaluate the parameters $s$ and $\mathcal F$ along the fluid-solid and fcc-bcc phase transitions in IPL and GCM systems we used the densities of the fluid and fcc phases, respectively. For the Yukawa system, the density gap between fcc and bcc phases as well as between fluid and solid phases was not resolved in Refs.~\cite{Hamaguchi,Hoy}. The values of $s_{\rm TP}$ and ${\mathcal F}_{\rm TP}$ employed are given in Table~\ref{Tab1}. The black solid curve corresponds to the universal melting equation ${\mathcal F}(s)\simeq 106 s^{2/3}$ ~\cite{Principle} in the regime of sufficiently soft interactions ($0.1\lesssim s\lesssim 1$)~\cite{Note2}.

Figure \ref{Fig2} demonstrates that the phase diagrams of the IPL, Yukawa, and GCM systems are topologically equivalent in the vicinity of the fcc-bcc-fluid triple point. The triple point separates the region where the fluid freezes directly into the fcc crystalline structure (regime of small $s$) from that where the intermediate bcc phase is present (regime of large $s$).
Another observation from Fig.~\ref{Fig2} is that the region of stability of the bcc phase in the $(s, {\mathcal F})$ plane widens when going from the IPL to Yukawa and further to GCM interaction. This suggests that the stability of the bcc phase is promoted by the short-range character of the interaction.


Let us now give several examples of a practical application of the obtained results. Recently, it has been suggested that the Lennard-Jones-type 7-6 potential, defined as $U(r)=A\epsilon[(\sigma/r)^7-(\sigma/r)^6]$ (where $A=7^7/6^6$), exhibits freezing into a stable bcc structure at sufficiently high temperatures~\cite{Sousa}. In particular, preliminary results from Ref.~\cite{Sousa} indicate that the fcc-bcc-fluid triple point temperature should be located somewhere in the interval $5.0 \lesssim T_* \lesssim 100$. Since the softness of the LJ-type 7-6 potential
tends to $s=0.143$ from below in the high-temperature limit, it is very likely that the triple point does exist. However, very weak dependence of $s$ on $\rho_*$ in this regime implies extreme sensitivity of the estimated triple point parameters to the input value of $s_{\rm TP}$. In this case, the criterion (\ref{criterion}) is not very useful. Taking the lower-limit value $s_{\rm TP}= 0.12$ suggested above, results in $\rho_*\simeq 3.4$. From the generic equation relating $T_*$ and $\rho_*$ of the $n-6$ LJ fluids at freezing  (Eq. (3) from Ref.~\cite{LJ-type}) we get $T_*\simeq 32.3$. This lower estimate of the triple point temperature falls in the interval predicted in ~\cite{Sousa}. Note that the 7-6 LJ potential is the only member of the $n$-6 LJ family which is expected to form a stable bcc structure.

Another model system displaying interplay between solid fcc and bcc phases is the system of particles interacting via the exp-6 potential of the form $U(r)=\epsilon\{\frac{6}{\alpha-6}\exp[\alpha(1-r/\sigma)]-\frac{\alpha}{\alpha-6}(\frac{\sigma}{r})^6\}$ beyond the hard-core at $r=r_0$, such that $U(r)$ exhibits maximum at $r_0$.  We concentrate here on the case $\alpha = 13$, which was considered in several studies. Still, the exact location of fcc-bcc-fluid triple point is a controversial issue. In Ref.~\cite{Belonoshko} triple point temperature has been estimated as $T_*\simeq 11$, while in a later study~\cite{SaijaPRB} a higher value $T_*\simeq 20$ has been reported. Applying the criterion (\ref{criterion}) yields for the triple point density and temperature: $\rho_*\simeq 4.4$ and $T_*\simeq 31.3$. Significant scattering in the triple point temperatures indicates that further studies are probably warranted to accurately locate the fcc-bcc-fluid triple point in this model.

One more potential, which is used to model the repulsive interactions in systems of soft particles, is the so-called Hertzian potential. It is defined as $U(r)=\epsilon(1-r/\sigma)^{5/2}$ for $r<\sigma$ and $U(r)=0$ otherwise. This potential is bounded (like GCM) and, additionally, has a finite interaction range. A detailed study of the phase diagram of Hertzian spheres has been performed by P\`{a}mies {\it et al}.~\cite{ Pamies}. A complicated picture with multiple re-entrant melting transitions along with first-order phase transitions between crystals with different symmetries has been reported. Nevertheless, the low density part of this phase diagram displays freezing upon compression either into a fcc or bcc crystal, the fcc-bcc phase transition, and the fcc-bcc-fluid triple point. Applying the criterion (\ref{criterion}) to the Hertz interaction results in the following estimate for the triple point density and temperature: $\rho_*\simeq 1.7$ and $T_*\simeq 4.6\times 10^{-3}$. These values are somewhat lower than the triple point parameters $\rho_*\simeq 1.96$ and $T_*\simeq 7.69\times 10^{-3}$ reported in Ref.~\cite{Pamies}, but still give an acceptable first guess, which can then be further improved by detailed numerical simulations.

Our last example is related to the Yoshida-Kamakura (YK) interaction law, $U(r)=\epsilon\exp[a(1-r/\sigma)-6(1-r/\sigma)^2\ln(r/\sigma)]$, where $a$ is a positive parameter controlling the softness of the repulsion. Phase diagram of the YK model with $a=2.1$ has been reported in Ref~\cite{SaijaPRE2009} and with $a=3.3$ in Ref.~\cite{PrestipinoJCP2010}. Again, the reported phase diagrams are quite complex, with multiple re-entrant regions and reach solid polymorphism. The low-density part, however, is marked by the presence of the fcc and bcc solid phases, phase transition between them, and the fcc-bcc-fluid triple point. The formulated criterion (\ref{criterion}) yields triple-point density and temperature  $\rho_*\simeq 0.24$ and $T_*\simeq 0.038$ for $a=2.1$, and $\rho_*\simeq 0.28$ and $T_*\simeq 0.033$ for $a=3.3$. These estimates are in fairly good agreement with the numerical results from Refs.~\cite{SaijaPRE2009} and \cite{PrestipinoJCP2010} (namely, $\rho_*\simeq 0.21$ and $T_*\simeq 0.040$ for $a=2.1$ and $\rho_*\simeq 0.25$ and $T_*\simeq 0.037$ for $a=3.3$\cite{SaijaPC}).

Thus, the existence and {\it approximate} location of the fcc-bcc-fluid triple point can be often predicted from the knowledge of the first two derivatives of the pair interaction potential, evaluated at the mean interparticle separation. Although the method is not expected to provide very high accuracy, its extreme simplicity can make it a useful tool for the preliminary order of magnitude estimates. In particular, it can be helpful when exploring the unique advantages of complex plasmas, associated with diversity and variability of interaction mechanisms between the macroparticles~\cite{FortovPR,Book,ChaudhuriSM,CPP}, in order to visit different regions of their phase diagram. Present results would be particularly relevant to situations when the possibilities to tune and design interactions are present. One of the possible methods that has been recently discussed~\cite{Komp} is based on the application of external electric fields of various polarizations. The physical idea is that the applied field induces ion flow, which generates wakes downstream from the particles, and thus change interparticle interactions (of course, the frequency should be much higher than that characterizing particle dynamics, so that the particles do not react to the field). Different anisotropic interaction classes as well as conventional isotropic interactions with short-range repulsion and long-range attraction can be obtained~\cite{Komp}.

In order to realize these active manipulations of the interparticle potential a ``PlasmaLab'' project has been initiated~\cite{PlasmaLab}. This is a possible followup of the PK-3 Plus and PK-4 laboratories~\cite{PK3+,PK4} for complex plasma investigations under microgravity conditions onboard the International Space Stations. Two possible plasma chambers are under active investigations. One chamber has a traditional cylindrical geometry with a flexible inner geometry and electrode system. The other has a spherical-like geometry with twelve independently controlled electrodes. Both chambers are expected to extend the plasma parameter space available for investigations and support interparticle force manipulations. In this context present results could be quite helpful for preliminary estimates of the experimental parameter regimes required to visit a desired phase state of a complex plasma under investigation.

\end{document}